\pdfoutput=1

\documentclass[twocolumn,showpacs,showkeys,amssymb,aps,nofootinbib,floatfix,superscriptaddress]{revtex4-1}

\usepackage{epsfig}
\usepackage{hyperref}
\usepackage{amssymb}
\usepackage{amsmath}

\DeclareMathOperator*{\E}{\mathbb{E}}

\begin{document} 

\title{\bf Statistical moments in superposition models and strongly intensive measures}

\author{Wojciech Broniowski}
\email{Wojciech.Broniowski@ifj.edu.pl}
\affiliation{The H. Niewodnicza\'nski Institute of Nuclear Physics, Polish Academy of Sciences, 31-342 Cracow, Poland}
\affiliation{Institute of Physics, Jan Kochanowski University, 25-406 Kielce, Poland}

\author{Adam Olszewski}
\email{Adam.Olszewski.fiz@gmail.com}
\affiliation{Institute of Physics, Jan Kochanowski University, 25-406 Kielce, Poland}

\begin{abstract}
First, we present a concise glossary of formulas for composition of standard, cumulant, 
factorial, and factorial cumulant moments in superposition (compound) models, where final
particles are created via independent emission from a collection of sources.
Explicit mathematical formulas for the composed moments are given to all orders. We discuss the
composition laws for various types of moments via the generating-function methods and list the
formulas for the unfolding of the unwanted fluctuations. Second, the technique is applied to the
difference of the scaled multiplicities of two particle types. This allows us for a systematic derivation
and a simple algebraic interpretation of the so-called strongly intensive fluctuation measures. With
the help of the formalism we obtain several new strongly intensive measures involving
higher-rank moments. The reviewed as well as the new results may be useful in investigations
of mechanisms of particle production and event-by-event fluctuations in high-energy nuclear and
hadronic collisions, and in particular in the search for signatures of the QCD phase transition at a
finite baryon density.
\end{abstract}

\date{4 April 2017}

\pacs{25.75.-q, 25.75Gz, 25.75.Ld}

\keywords{ultra-relativistic nuclear collisions, cumulants, factorial moments, superposition models, event-by-event fluctuations, QCD critical point, strongly intensive measures}

\maketitle

\section{Introduction}

In recent years, intense activity has been focused on possible detection of critical phenomena in QCD. A basic tool of these investigations are the statistical moments 
of the multiplicities (or, in general, one-body observables such as momenta or charges) of the produced and experimentally detected particles, 
with the premise that the large fluctuations linked to critical phenomena will remain manifest in the 
experimentally detected particle distributions. The primary objects in these studies are the cumulant~\cite{Fisher} and factorial moments~\cite{Riordan}, vastly used in statistical 
studies in various domains of science.  
The very long history of applications of cumulants to particle physics includes, to mention a few,  the studies of intermittency
via factorial moments~\cite{Bialas:1985jb}, analysis of fluctuations in gluodynamics~\cite{Dremin:1993sd}
or investigations of ratios of factorial to cumulant moments~\cite{PhysRevC.58.1720,*Rybczynski:1999pa}.
In the field of ultra-relativistic heavy-ion collisions, the cumulants in the azimuthal angle have become
a standard tool in studies of the harmonic flow~\cite{Borghini:2000sa,Bilandzic:2010jr}.
General mathematical features and the combinatorial interpretation of cumulants, including the multivariate case, 
have been recently reviewed in~\cite{DiFrancesco:2016srj}, with 
a stress on applications to harmonic flow. 

From the practical point of view, important applications concern the unfolding of uninteresting fluctuations, 
as studied, e.g., in~\cite{Fu:2003yf,Bialas:2003gq,Ling:2015yau,Nonaka:2017kko}
Similar goals were addressed in~\cite{Gazdzicki:1992ri}, or more recently in~\cite{Gorenstein:2011vq,Gazdzicki:2013ana} by means of 
the so-called {\em strongly intensive}
measures. 

The focus of our study are the superpositions of distributions, also known in statistics as {\em compound} distributions (see, e.g.,~\cite{Bean}):
The initial {\em sources}\footnote{The nature of the source depends on a particular model. The operational definition is that it 
emits particles independent from other sources, whereas the distribution of sources themselves may in general be correlated.} emit independently particles with a certain distribution, which results  
in a final (measured) distribution of particles. More superposition 
steps~\cite{Olszewski:2013qwa,Olszewski:2015xba} may be needed in a realistic description of the production process, involving a hydrodynamic stage in the 
intermediate step.

The purpose of this paper is twofold: First, we bring standard techniques and results to the attention of practitioners in the field 
of particle/heavy-ion physics, where the material presented 
in a form of a concise glossary can be useful.
In particular, we present exact relations between various kinds of moments (standard, cumulant, factorial, and factorial cumulant) to all 
orders, which 
involve Bell polynomials or the Stirling numbers (Sec.~\ref{sec:genfun}). 

Next, we present explicit formulas for the composed moments in superposition models to all orders. 
The structure of  the composition laws for various kinds of moments follows directly from  the composition properties of the corresponding generating functions, 
which are basic objects in our derivations.
These composition properties are particularly simple for certain combinations of types of moments (Sec.~\ref{sec:onevar}). 
We also consider the inverse problem, where one infers the distribution of sources from the known distribution of particles and the overlaid distribution. 
This unfolding procedure eliminates the unwanted/trivial fluctuations. Two important cases are the unfolding of the detector 
efficiency~\cite{Whitmore:1976ip,Foa:1975eu}, typically 
modeled with a superposed Bernoulli trial, and the removal of thermal fluctuations, which lead to an overlaid Poisson distribution. 

Second (Sec.~\ref{sec:strong} and \ref{sec:dp}), we present a novel systematic way to derive the strongly intensive measures~\cite{Gazdzicki:1992ri}, which follows from the consideration of cumulants for the difference 
of multiplicities (or momenta or charges) of two types of particles. That way we are able to generalize the results of~\cite{Gorenstein:2011vq,Gazdzicki:2013ana} and obtain new relations involving higher-rank 
moments. The relations involve {\em identical in form} combinations of moments for the particles produced from a source and for the final particles, 
similarly to the case of the rank-2 formulas from~\cite{Gorenstein:2011vq}

Our results can be useful in the actively pursued investigations of mechanisms of particle production and event-by-event fluctuations, in particular in the 
search of the QCD phase 
transition at finite baryon density~\cite{Stephanov:1998dy,Stephanov:1999zu,Stephanov:2008qz,Cheng:2008zh,Gupta:2009mu,Athanasiou:2010kw,%
Gupta:2011wh,Ling:2015yau,Begun:2016cva,Begun:2016sop,Braun-Munzinger:2016miz,Braun-Munzinger:2016yjz}. 
An active search program is on the way by the NA61 Collaboration~\cite{Gazdzicki:2015ska}.
Other aspects of correlations and fluctuations in relativistic heavy-ion collisions are reviewed in~\cite{Koch:2008ia}.
Experimentally obtained cumulants of the net proton distributions~\cite{Aggarwal:2010wy} and the 
net charge distributions~\cite{Adamczyk:2014fia,Adare:2015aqk} have been recently analyzed in~\cite{Bzdak:2012ab,Bzdak:2012an,Bzdak:2016qdc,Almasi:2017bhq}.
A review of up-to-date lattice results can be fond in~\cite{Karsch:2016yzt}, whereas a study in the  UrQMD model of the STAR experimental data
has been carried out in~\cite{Xu:2016qjd,He:2017zpg}.

Sections~\ref{sec:genfun}-\ref{sec:more} have mostly an introductory character, preparing grounds for Sec.~\ref{sec:strong} and \ref{sec:dp}, where new results for the 
strongly intensive fluctuation  measures are derived.
Some potentially useful relations between moments in the superposition models are given in the Appendices. 

\section{Generating functions for various types of moments\label{sec:genfun}}

To establish the framework and notation, we begin with recalling 
the definitions of generating functions for various kinds of statistical moments and their inter-relations, 
which in turn lead to linear relations between various kinds of moments themselves.

The generating function for the standard moments of a random variable $X$ is defined as\footnote{Throughout the paper we adopt the convention 
that a superscript in various quantities indicates the given random variable.} 
\begin{eqnarray}
M^X(t)\equiv\E_X e^{tX}=1+ \sum_{i=1}^\infty \mu'^X_i \frac{t^i}{i!}, \label{eq:mu}
\end{eqnarray}
where $\E_X$ denotes the operator of averaging over $X$ and $ \mu'^X_i =  \E_X X^i$ are the standard (centered about the origin) moments of $X$.
The generating function for the {\em cumulant} moments  $\kappa^X_i$~\cite{Fisher}  is defined as 
\begin{equation}
K^X(t)\equiv \log M^X(t) = \sum_{i=1}^\infty \kappa^X_i \frac{t^i}{i!}. \label{eq:K}
\end{equation}
The relation between the standard and cumulant moments is as follows:
\begin{eqnarray}
&&\mu'^X_m=\sum_{k=1}^m B_{m,k}(\kappa^X_1,\dots,\kappa^X_{m-k+1}), \nonumber \\
&&\kappa^X_m=\sum_{k=1}^m  (-1)^k (k-1)! B_{m,k}(\mu'^X_1,\dots,\mu'^X_{m-k+1}), \label{eq:bell1}
\end{eqnarray}
where $B_{m,k}(x_1,\dots,x_{m-k+1})$ denote the partial (a.k.a. incomplete) exponential Bell polynomials~\cite{wiki:Bell}.
The combinatorial meaning of these polynomials lies in the encoding of the information on set partitions: the coefficients of the 
subsequent monomials in  $B_{m,k}$ are equal to the number of partitions of an $m$-element set into $k$ non-empty subsets. 
More precisely, the coefficient of the monomial $x_{i_1}^{p_1} \dots x_{i_s}^{p_s}$ is equal to the number of partitions into 
subsets with $i_1, \dots, i_s$ elements, where the subsets occur $p_1, \dots, p_s$ times. For instance, 
$B_{4,2}(x_1,x_2,x_3) = 4 x_1 x_3+3 x_2^2$, showing that we can partition a 4-element set into subsets of one- and three elements in 4 ways, and 
into two subsets of 2 elements in 3 ways. That way we can interpret the upper Eq.~(\ref{eq:bell1}) as 
a decomposition of $\mu'^X_m$ into ``connected'' components  $\kappa^X_i$.

The generating function for the {\em central} moments $\mu^X_i$ is
\begin{eqnarray}
C^X(t)=e^{-\mu^X t} M^X(t) = 1+ \sum_{i=1}^\infty \mu^X_i \frac{t^i}{i!},
\end{eqnarray}
with $\mu^X=\E_X X$ denoting the average. The relation with the cumulant moments is
\begin{eqnarray}
&&\mu^X_m=\sum_{k=1}^m B_{m,k}(0,\kappa^X_2,\dots,\kappa^X_{m-k+1}), \label{eq:bellc}\\
&&\kappa^X_m=\sum_{k=1}^m  (-1)^{k+1} (k-1)! B_{m,k}(0,\mu^X_2,\dots,\mu^X_{m-k+1}), \nonumber \\
\end{eqnarray}
with $m\ge 2$.

The generating function for the {\em factorial} moments 
\begin{eqnarray}
f_i^X=\E_X X(X-1)\dots(X+1-i) \label{eq:factmom}
\end{eqnarray}
is
\begin{eqnarray}
F^X(t)= \E_X (1+t)^X = 1+ \sum_{i=1}^\infty f^X_i \frac{t^i}{i!}, \label{eq:Fx}
\end{eqnarray}
having the interpretation of the average number, average number of pairs, average number of triples, {\em etc.}
From definition, it is related to the generating function for the cumulant moments via the change of variables:
\begin{eqnarray}
F^X(t)=M^X [\log(1+t)] = e^{K^X[\log(1+t)]}. \label{eq:FM}
\end{eqnarray}
Finally, the generating function for the {\em factorial cumulant} moments, $\kappa'^X_m$, is defined as 
\begin{eqnarray}
G^X(t) \equiv K^X[\log(1+t)]=\log \left [ F^X(t) \right ]= \sum_{i=1}^\infty \kappa'^X_i \frac{t^i}{i!}. \nonumber \\ 
\label{eq:GF}
\end{eqnarray}
The relation of factorial cumulant moments to factorial moments is fully analogous to the relation of cumulant moments to standard moments,  
with the same combinatorial interpretation as discussed above.

\begin{figure}[tb]
\begin{center}
\includegraphics[width=.221\textwidth]{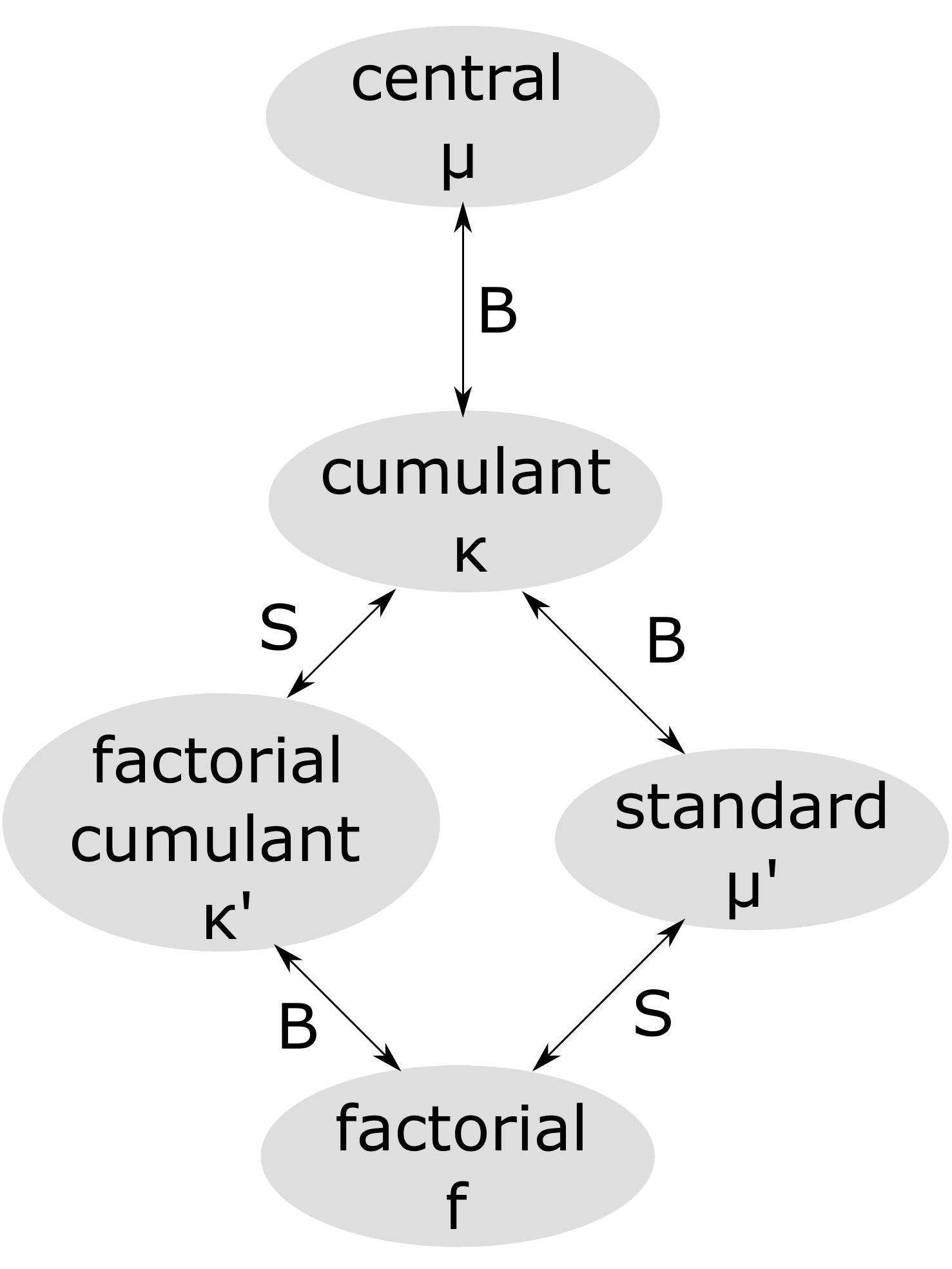}
\end{center}
\vspace{-3mm}
\caption{Summary of the transformations between various kinds of moments. $B$ denotes the transformation with 
partial Bell polynomials (as in Eq.~(\ref{eq:bellc},\ref{eq:bell2}), and $S$ denotes the transformation with 
the Stirling numbers (as in Eq.~(\ref{eq:s12}). \label{fig:tr}}
\end{figure}

As the functions $F^X$ and $M^X$ in Eq.~(\ref{eq:FM}) and the functions $G^X$ and $K^X$ in Eq.~(\ref{eq:GF}) are related through a 
simple change of variables $t \leftrightarrow \log(1+t)$, the 
corresponding moments are linked with a linear transformation involving the Stirling numbers as the coefficients.   
Explicitly, 
\begin{eqnarray}
f^X_m&=&\sum_{k=1}^m s(m,k)  \mu'^X_m, \;\;\; \mu'^X_m=\sum_{k=1}^m S(m,k)  f^X_m,  \nonumber \\
\kappa'^X_m&=&\sum_{k=1}^m s(m,k)  \kappa^X_m, \;\;\; \kappa^X_m=\sum_{k=1}^m S(m,k)  \kappa'^X_m,  \label{eq:s12} 
\end{eqnarray}
where $s(m,k)$ are the signed Stirling numbers of the first kind (cycle numbers), and 
$S(m,k)$ are the Stirling numbers of the second kind (partition numbers). 
Finally, we note that the relation between $G^X$ and $F^X$  in Eq.~(\ref{eq:GF}) has the same form as
the relation between $K^X$ and $M^X$ in Eq.~(\ref{eq:K}), therefore  
\begin{eqnarray}
&&f^X_m=\sum_{k=1}^m B_{m,k}(\kappa'^X_1,\dots,\kappa'^X_{m-k+1}), \label{eq:bell2} \\
&&\kappa'^X_m=\sum_{k=1}^m  (-1)^{k+1} (k-1)! B_{m,k}(f^X_1,\dots,f^X_{m-k+1}), \nonumber
\end{eqnarray}
The scheme of relations between various kinds of moments are summarized in Fig.~\ref{fig:tr}, where the labels on the arrows indicate
the transformations via the partial Bell polynomials ($B$), or the Stirling numbers ($S$). 

A remark at this point is that 
it is much a matter of convenience and convention which type of moments to choose in a given analysis, as they are all 
directly related via linear transformations. Nevertheless, physical argument may indicate advantages of a certain type of moments 
in a given analysis (cf. Sec.~\ref{sec:special}).

\section{Generating functions in the superposition model \label{sec:onevar}}

In {\em superposition} models\footnote{In other domains 
of application of statistics, these models are frequently referred to as {\em compound} models~\cite{Bean}.} of hadron production, 
the number of particles $N$, as registered in the experiment, is composed from independent production from $S$ sources; the $j$-th source produces $n_j$ particles, {\em i.e.},
\begin{eqnarray}
N=\sum_{j=1}^S n_j.  \label{eq:N}
\end{eqnarray}
The variables $n_j$ are random, and so is the number of sources $S$. All the variables $n_j$ and the multiplicity of sources $S$  are, by assumption 
on the production mechanism, {\em independent} from 
one another, which is crucial in the following derivation of the composition formulas. 
The number of sources $S$ fluctuates event-by-event, hence the multiplicity distribution of the finally produced particles reflects these fluctuations, 
as well as the fluctuations in the random variables $n_j$.

For simplicity, we assume that the production from each source is the same, 
hence all $n_j$ have the same distribution and a common cumulant generating function $K^n(t)$. We start with the case of one type 
of sources, with the straightforward generalization to more types of sources presented in Sec.~\ref{sec:more}.

One should bear in mind that what we call in this paper ``multiplicity'' may in fact refer to any additive one-body characteristics, such as charge of momentum. For instance, $N$ could 
stand for the total charge in the event, and $n_i$ for the charge of particles produced by the source $i$.

We remark here that our notion of the {\em source}, whereby emission from different sources is by definition independent from each other, is 
quite restrictive from the point of view of the global conservation laws of the energy-momentum or
charges (for reviews of the conservation laws effects see, e.g.,~\cite{Jeon:2003gk,Mrowczynski:2009wk} 
and for the influence on the strongly intensive measures see~\cite{Zaranek:2001di,Mrowczynski:2001mm}). 
Naive imposition of such global conservation constraints on the momenta or charges of the produced particles
would necessarily correlate production from different sources, which would be at odds with the basic assumption.
The issue may be resolved by introducing various types of sources (labeled, for instance, by the value of the charge) and 
keeping track of the {\em local} conservation laws at the level of the production from a given source.
We return to this issue at the end of the Conclusion section, where we indicate how to introduce more types of sources and the 
local conservation laws, generalizing the approach.

Substitution of Eq.~(\ref{eq:N}) into  Eq.~(\ref{eq:K}) yields immediately
\begin{eqnarray}
e^{K^N(t)} &=&  \E_N e^{tN} =  \E_{S,n_1,\dots,n_S} e^{\sum_{j=1}^S n_j}\label{eq:main} \\ &=& \E_{S,n_1,\dots,n_S} \prod_{j=1}^S e^{t n_j} =  \E_S  e^{S K^n(t)} 
= e^{K^S[K^n(t)]}. \nonumber
\end{eqnarray}
In the third equality we have used the fact that $n_i$ and $n_j$ are not correlated for $i \neq j$, and in the 
last equality we have used the definition of the cumulant generating function for the distribution of the number of sources $S$, which in the formula takes
the argument $K^n(t)$. Thus we have arrived at a known fact for the compound models, namely, 
that the independent superposition of distributions leads to the composition of the corresponding 
cumulant generating functions~\cite{Bean}, 
\begin{eqnarray}
K^N(t)=K^S \left [ K^n(t) \right ]. \label{eq:cumcomp}
\end{eqnarray}

In addition to the composition law of Eq.~(\ref{eq:cumcomp}), one may straightforwardly derive additional relations.
For the standard moment generating functions the following composition law follows:
\begin{eqnarray}
M^N(t)=M^S\left( K^n(t)\right)=M^S\left ( \log \left [ M^n(t) \right ] \right ),  \label{eq:stdcomp}
\end{eqnarray}
whereas for the central moment generating function we find
\begin{eqnarray}
 C^N(t)&=&\left[ C^n(t)\right]^{\mu^S} C^S\left (K^n(t) \right ) \nonumber \\ &=& \left[ C^n(t)\right]^{\mu^S} C^S\left ( \log[C^n(t)]+\mu^n t\right ).  \label{eq:cencomp}
\end{eqnarray}
For the factorial moment generating function one arrives, after a short calculation, at the composition law
\begin{eqnarray}
F^N(t)-1=F^S \left [ F^n(t)-1 \right ] -1, \label{eq:Fcomp}
\end{eqnarray}
which takes the same form as Eq.~(\ref{eq:cumcomp}). For that reason all the general
statements concerning the standard and cumulant moments also hold for the factorial 
and factorial cumulant moments. 
Finally, for the factorial cumulant moment generating function we find
\begin{eqnarray}
G^N(t)=G^S \left [ F^n(t) -1 \right ]=G^S \left [ e^{G^n(t)}-1 \right ]. \label{eq:Gcomp}
\end{eqnarray}
Additional formulas are obtained via the replacement $t \to \log(1+t)$ in Eq.~(\ref{eq:cumcomp},\ref{eq:stdcomp}), which yields 
\begin{eqnarray}
G^N(t)=K^S \left [ G^n(t) \right ], \;\;\;  F^N(t)= M^S \left [ G^n(t) \right ]. \label{eq:GKcomp}
\end{eqnarray}

The obtained composition laws are collected in Table~\ref{tab:comp}, which is central for this part of our paper. 
We list formulas both in the autonomous form, {\em i.e.}, involving the
generating functions for moments of one type only, as well as in the form of compositions of the 
generating functions of different types, which may also be useful in some applications.

\begin{table}[tb]
\caption{Composition laws for generating functions in the superposition model. 
We list the forms which are autonomous, {\em i.e.}, involve the generating functions of the same type, 
and the forms which correspond to compositions of functions of various types.\label{tab:comp}}
\begin{center}
\bgroup
\def\arraystretch{1.5}
\begin{tabular}{|ll|}
\hline
type of moments & composition formula \\
\hline 
central               & $C^N(t)=\left[ C^n(t)\right]^{\mu^S} C^S\left [ \log C^n(t)+\mu^n t\right ]$  \\
                          & ~~~~~~~~~$=\left[ C^n(t)\right]^{\mu^S} C^S\left [ K^n(t) \right ]$ \\
\hline 
standard            & $M^N(t) =M^S\left [ \log  M^n(t)  \right ]=M^S\left [ K^n(t) \right] $ \\ \hline
cumulant           & $K^N(t)=K^S[K^n(t)]$ \\ \hline
factorial             & $F^N(t)-1=F^S \left [ F^n(t) -1 \right ] -1$ \\
                          & $F^N(t)= M^S \left [ G^n(t) \right ]$ \\ \hline
factorial             & $G^N(t)=G^S [ e^{G^n(t)}-1]$ \\
cumulant           & $~~=G^S \left [ F^n(t)-1 \right ]= K^S \left [ G^n(t) \right ]$                      \\  \hline
\end{tabular}
\egroup
\end{center}
\end{table}

\section{Moments in the superposition model\label{sec:momsup}}

We note from the formulas in Table~\ref{tab:comp} that certain compositions of generating functions for various types of moments have a simple 
form of a composite function:  $M^N(t) =M^S\left [ K^n(t) \right] $, $K^N(t)=K^S[K^n(t)]$, {\em etc.} Note that the formulas may involve various types of moments. 
When the composition has a generic form 
\begin{eqnarray}
P^N(t)=Q^S[R^n(t)], \label{eq:gengen}
\end{eqnarray}
where $P$, $Q$, $R$ stands for $M$, $K$, $F-1$, or $G$,
we may obtain the corresponding moments in a particularly simple manner:\footnote{Of course, we may always obtain the composition law for the moments 
via the Maclaurin expansion of the generation function of any form.}. Namely, one can use the Fa\`a di Bruno's formula
for the $n$-th derivative of a composite function to arrive at the formula for the corresponding moments (denoted with the same generic symbol as the generating 
functions):
\begin{eqnarray}
P_m=\sum_{k=1}^m Q_k B_{m,k}(R_1,\dots,R_{m-k+1}),  \label{eq:bell}
\end{eqnarray}
where again we encounter the exponential partial Bell polynomials discussed in Sec.~\ref{sec:genfun}.
We have denoted the moments with the same generic symbol as for the generation function in Eq.~(\ref{eq:gengen}), for instance, the 
moments corresponding to $P^N(t)$ are $P_m$.
Explicit formulas for the first few values of $m$ are listed in App.~\ref{app:explicit}.

\begin{figure}[tb]
\begin{center}
\includegraphics[width=.195\textwidth]{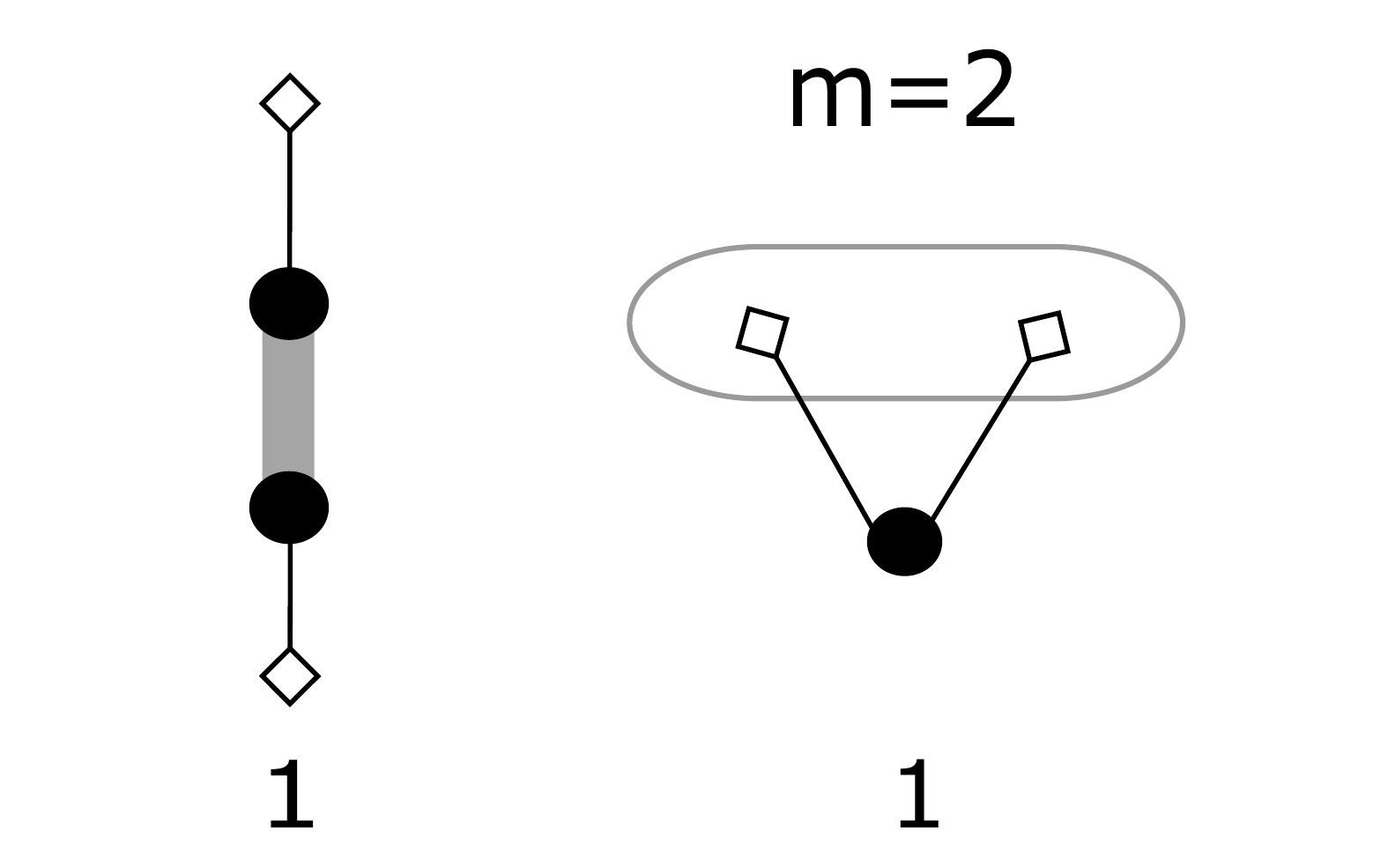} \\ \vspace{7mm}
\includegraphics[width=.32\textwidth]{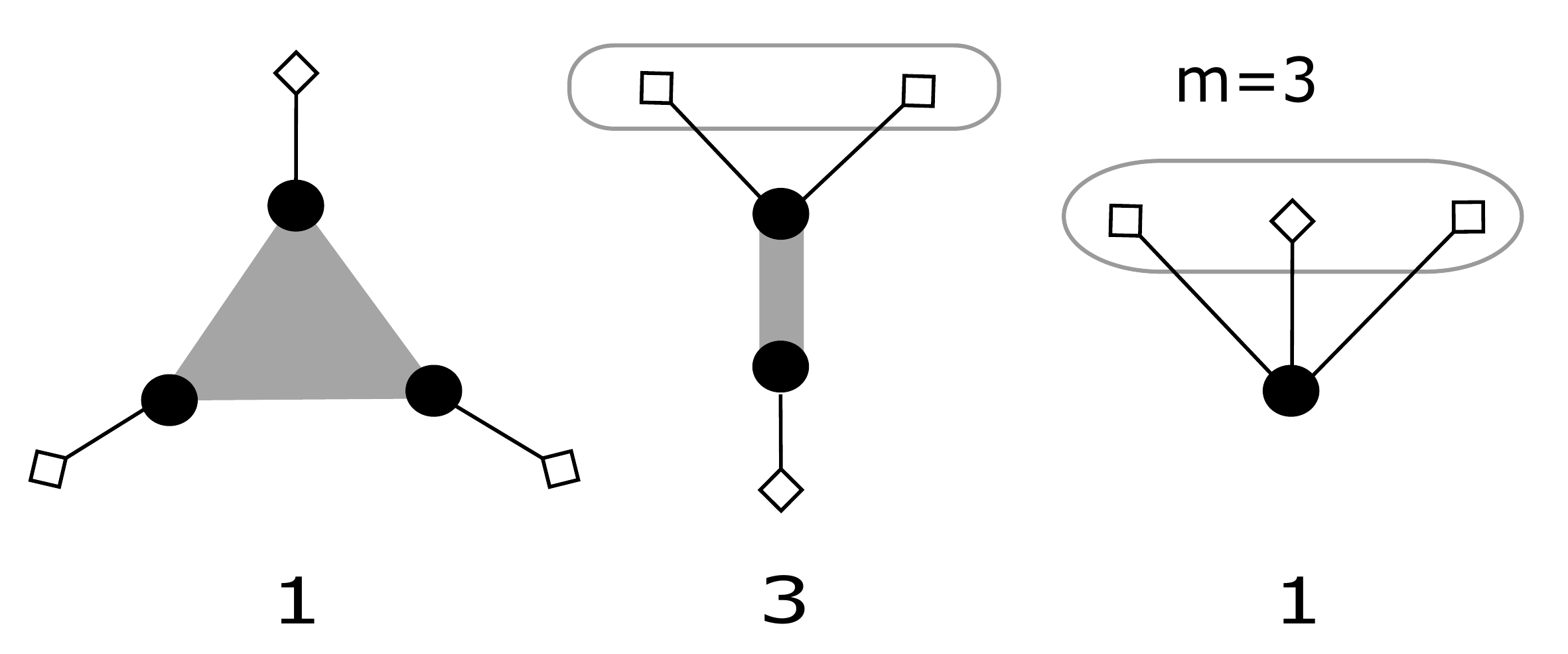}\\ \vspace{7mm}
 \includegraphics[width=.5\textwidth]{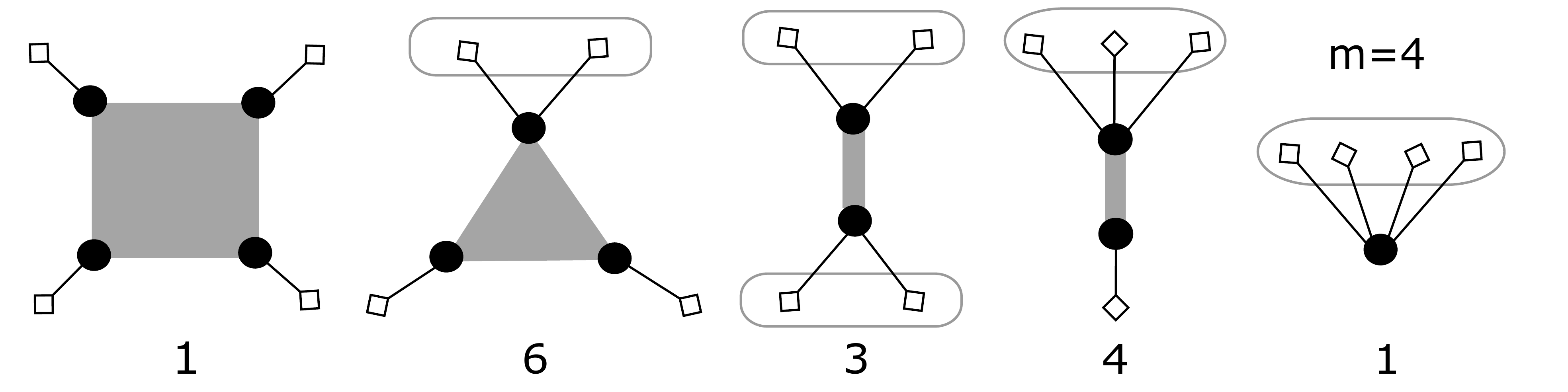}
\end{center}
\caption{Graphical interpretation of the composition formulas for the moments in the superposition model. The dark blobs connected with shaded regions 
correspond to the sources, and the empty squares to the produced particles. A particle is linked with the source which produced it. Ovals encircling particles 
indicate the connected correlations. See text for more details. \label{fig:int}}
\end{figure}

The probabilistic interpretation of Eq.~(\ref{eq:bell}), directly related to the superposition model, is visualized in Fig.~\ref{fig:int}. 
The blobs connected with shaded regions indicate the sources which are correlated in a connected way. The empty squares stand 
for $m$ produced particles, and each particle is linked with its own source. Ovals encircling particles 
indicate the connected correlations. The labels below the diagrams, which are the corresponding coefficients in the partial Bell polynomials, are interpreted 
as the number of inequivalent labeling of the squares in the graphs, with the blobs remaining unlabeled. For instance, in the second diagram 
of the bottom row (for $m=4$) we find $6$
possibilities, {\em i.e.}, the number of ways for selecting a pair from a four-element set. 

One is often interested in the inverse problem, where we know the moments of the $N$ and $n$ distributions, and wish to infer the corresponding moments of the source number $S$.
This is the {\em unfolding} procedure for the moments.
In that case we rewrite the composition law (\ref{eq:gengen}) as
\begin{eqnarray}
Q^S(u)=P^N[(R^n)^{-1}(u)], \label{eq:invcomp}
\end{eqnarray}
where
\begin{eqnarray}
(R^n)^{-1}(u)=\sum_{j=1}^\infty \lambda_j \frac{u^j}{j!}
\end{eqnarray}
defines the inverse generating function for the overlaid distribution. Hence, in analogy to Eq.~(\ref{eq:bell}),
\begin{eqnarray}
Q_m=\sum_{k=1}^m P_k B_{m,k}(\lambda_1,\dots,\lambda_{m-k+1}). \label{eq:ks}
\end{eqnarray}
Through the Lagrange inversion formula of the Maclaurin series (starting with a linear term) we 
can express the coefficients $\lambda_j$ through the moments $R_i$ as follows:
\begin{eqnarray}
&& \lambda_1=\frac{1}{R_1}, \label{eq:lag} \\
&& \lambda_j=\frac{1}{(R_1)^j} \sum_{k=1}^{j-1} (-1)^k \frac{(j+k-1)!}{(j-1)!} 
 B_{j-1,k}(\hat R_1,\dots,\hat R_{j-k}), \nonumber
\end{eqnarray}
for $j \ge 2$, where the scaled moments of $n$ are
\begin{eqnarray}
\hat  R_i=\frac{R_{i+1}}{ (i+1)R_1}.
\end{eqnarray}
The combination of Eqs.~(\ref{eq:ks}) and (\ref{eq:lag}) yields the inverse composition formulas for the
moments of the sources.\footnote{Alternatively, one 
may explicitly solve the triangular linear equation set (\ref{eq:bell}) or (\ref{app:one}) for the source cumulant coefficients $Q_m$ up to the desired order.} 
Explicit expressions for the first few values of $m$ in Eq.~(\ref{eq:ks}) are presented in Eq.~(\ref{app:two}) in App.~\ref{app:explicit}.

\section{Case of special overlaid distributions \label{sec:special}}

There are two physically relevant cases where the composition laws assume a very simple form, because one of the 
types of the generating function is linear in $t$. The first case occurs when the detector efficiency is modeled with a 
Bernoulli trial, with $p$ denoting success of the observation of a particle, and $q=1-p$ failure.
For the Bernoulli trial the simplest generating function is for the factorial moments, $F^n(t)-1=p t$. Then 
we find immediately from  Eq.~(\ref{eq:Fx},\ref{eq:Fcomp}) and Eq.~(\ref{eq:GF},\ref{eq:Gcomp}) that
\begin{eqnarray}
f^N_m=p^m f^S_m, \;\;\; \kappa'^N_m=p^m \kappa'^S_m,  \;\;\; ({\rm Bernoulli}) \label{eq:sc1}
\end{eqnarray}
which means a uniform scaling of the factorial and the factorial cumulant moments with powers of $p$.

The other important case is the Poisson distribution, for which
the simplest is the factorial cumulant generating function 
$G(t)=\beta t$, with $\beta$ denoting the mean. This case is encountered in modeling statistical hadronization 
from thermal sources.\footnote{It may still be followed with a Bernoulli trial of the detector efficiency, as a composition of the Poisson distribution with a Bernoulli trial is again a Poisson distribution 
with the mean equal to $p \beta$.} Then we find the simple relations
\begin{eqnarray}
f^N_m=\beta^m \mu'^S_m, \;\;\; \kappa'^N_m=\beta^m \kappa^S_m,  \;\;\; ({\rm Poisson})  \label{eq:sc2}
\end{eqnarray}
linking the factorial moments of particles with the standard moments of sources, 
and the factorial cumulant moments of particles with the cumulant moments of sources. The scale coefficients $p$ or $\beta$ in Eq.~(\ref{eq:sc1},\ref{eq:sc2}) disappear when 
appropriate scale-less ratios of the moments are considered, for instance $\kappa'^N_4/(\kappa'^N_2)^2=\kappa^S_4/(\kappa^S_2)^2$ for the Poisson case.

Other popular distributions, such as the binomial distribution, the Gamma distribution, of the negative binomial distribution, do not lead to composition laws as simple as 
Eq.~(\ref{eq:sc1}) or (\ref{eq:sc2}), and in such cases one needs to use the general composition laws spelled out in  App.~\ref{app:explicit}.

Multiplicity-dependent and non-Bernoulli trial efficiency corrections were considered in~\cite{Bzdak:2016qdc}. The case of the non-Bernoulli trial corrections may be 
analyzed according to the general formulas (\ref{app:one}). Note that a strong sensitivity to the detector features advocated in~~\cite{Bzdak:2016qdc} may be 
attributed to large numerical coefficients appearing in Eqs.~ (\ref{app:one}), in particular for the higher-rank moments.

For the case where only two first moments of the overlaid distribution, $R_1$ and $R_2$, are nonzero, such as for instance in the case of the normal distribution with 
$K^n(t)=\mu t +\sigma^2 t^2/2$, we find the following composition formulas:
\begin{eqnarray}
P_n &=& \sum_{k=0}^{\left\lfloor \frac{n}{2}\right\rfloor } B(n,n-k)   R_1^{n-2 k} R_2^k  Q_{n-k}, \label{eq:normal} \\
Q_n &=& \sum_{k=0}^{n-1} b(n,n-k) (n-k)! R_1^{-k+n-1}  R_2^k P_{n-k}, \nonumber
\end{eqnarray}
where $B(n,k)$ and $b(n,k)$ are the Bessel numbers of the second and first kind~\cite{wiki:Bessel}, respectively. They are equal to 
\begin{eqnarray}
B(n,k) &=& \frac{2^{k-n} n!}{(2 k-n)! (n-k)!},  \nonumber \\  b(n,k) &=& \frac{\left(-\frac{1}{2}\right)^{n-k} (2 n-k-1)!}{(k-1)! (n-k)!}. \label{eq:Bessel}
\end{eqnarray}

If we wish to unfold other types of distributions, then the relations are more complicated than in Eqs.~(\ref{eq:sc1},\ref{eq:sc2}), 
keeping the generic triangular form of Eq.~(\ref{app:two}).
This is for instance the case of the $\Gamma$ distribution or the 
negative binomial distribution, frequently used to model 
the early production from the initial sources. In that case we can explicitly use the cumulant moment generating functions 
of the form
\begin{eqnarray}
K_\Gamma(t) = -a \log(1 - m t/a)
\end{eqnarray}
or 
\begin{eqnarray}
K_{\rm neg.~bin.}(t)=-n \log[(1 - q e^t)/(1 - q)]
\end{eqnarray}
(or the corresponding generating functions for other types of moments) 
in the composition formulas, and derive the appropriate relations via the Maclaurin expansion. 
Of course, one can alternatively use the 
general composition formulas of App.~\ref{app:explicit}.

\section{More kinds of sources\label{sec:more}}

In various models of particle production one may distinguish more kinds of sources. Examples are the wounded nucleon~\cite{Bialas:1976ed,*Bialas:2008zza} or wounded 
quark~\cite{Bialas:1977en,*Anisovich:1977av} models, where we have emitting sources (wounded objects) associated to the two colliding nuclei, $A$ or $B$. Another case 
occurs in considering the rapidity bins in studies of the longitudinal correlations, or, in general, separated bins in the kinematic space. 
In that situation, if production of particles in different  bins is independent 
from each other, one may formally treat the bins as (possibly correlated) sources. We present in detail the extension of the 
formalism to the case of two sources, as a generalization 
to more sources is obvious. 
We note that we now enter the domain of multivariate moments and cumulants~\cite{Fisher}, which have a similar combinatorial 
interpretation as the univariate case discussed in the previous sections.  

Let the two kinds of sources be denoted as $A$ and $B$. These sources are, in general, not independent of each other (they may be correlated), but as before the particles 
produced from {\em different} sources are independent from one another, and also independent of the multiplicity of the sources, denoted as  $S_A$ and $S_B$.
The cumulant generating function for two variates (the total number of produced particles of type $A$ and $B$),
\begin{eqnarray}
N_A=\sum_{j=1}^{S_A} n_j, \;\; N_B=\sum_{j=1}^{S_B} n_j,  \label{eq:NAB} 
\end{eqnarray}
is defined as
\begin{eqnarray}
{K^{N_A,N_B}(t_A,t_B)} &=& \log \left [ \E_{N_A,N_B} e^{t_A N_A+t_B N_B} \right ]   \nonumber \\ 
&=& \sum_{i,j=0,\,i+j>0}^\infty \frac{\kappa^{N_A,N_B}_{i,j}}{i! j!}t_A^i t_B^j ,   \label{eq:main2}
\end{eqnarray}
and similarly for the case of the sources $S_A$ and $S_B$.
Note that the sum in Eq.~(\ref{eq:main2}) involves also the terms with $\kappa^{N_A,N_B}_{i,0}=\kappa^{N_A}_{i}$ and $\kappa^{N_A,N_B}_{0,j}=\kappa^{N_B}_{j}$.
Generalizing the derivation of Sec.~\ref{sec:onevar} yields the relation
\begin{eqnarray}
e^{K^{N_A,N_B}(t_A,t_B)}&=&\E_{N_A,N_B}e^{t_A N_A+t_B N_B} \nonumber \\ &=&\E_{S_A,S_B} e^{S_A K^{n_A}(t_A)+S_B K^{n_B}(t_B)} 
\nonumber \\ &=&e^{K^{S_A,S_B}[K^{n_A}(t_A),K^{n_B}(t_B)]}, \label{eq:twovar}
\end{eqnarray}
hence 
\begin{eqnarray}
K^{N_A,N_B}(t_A,t_B)=K^{S_A,S_B}[K^{n_A}(t_A),K^{n_B}(t_B)].
\end{eqnarray}

We may now repeat the steps of Sec.~\ref{sec:momsup} to arrive at more generic composition laws between the generating 
functions of various kinds, 
in analogy to Table~\ref{tab:comp}:
\begin{eqnarray}
P^{N_A,N_B}(t_A,t_B)=Q^{S_A,S_B}[R^{n_A}(t_A),R^{n_B}(t_B)].
\end{eqnarray}
For the inverse problem 
\begin{eqnarray}
Q^{S_A,S_B}(u_A,u_B)=P^{N_A,N_B}[(R^{n_A})^{-1}(u_A),(R^{n_B})^{-1}(u_B)]. \nonumber \\  \label{eq:invprob2}
\end{eqnarray}
For the corresponding moments we find
\begin{eqnarray}
P_{m,n}=\sum_{k=0}^m {\sum_{l=0}^n}' Q_{k,l} && B_{m,k}(R_{A,1},\dots,R_{A,m-k+1}) \nonumber \\ && \times B_{n,l}(R_{B,1},\dots,R_{B,n-l+1}), \nonumber \\
Q_{m,n}=\sum_{k=0}^m {\sum_{l=0}^n}' P_{k,l} && B_{m,k}(\lambda_{A,1},\dots,\lambda_{A,m-k+1}) \nonumber \\ && \times B_{n,l}(\lambda_{B,1},\dots,\lambda_{B,n-l+1}), 
\nonumber \\ \label{eq:bellbis}
\end{eqnarray}
where an obvious generalization of the notation of Sec.~\ref{sec:momsup}  has been used, and the prime in the summation symbol indicates that the term $k=l=0$ is avoided.
Generalization of the above formulas to the case with more particle types is immediate, with 
\begin{eqnarray}
&& P^{N_A,N_B,N_C,\dots}(t_A,t_B,t_C,\dots)  \\ && ~~ =Q^{S_A,S_B,S_C}[R^{n_A}(t_A),R^{n_B}(t_B),R^{n_C}(t_C),\dots], \nonumber
\end{eqnarray}
and similarly for the inverse relation and the moments.

\section{More kinds of particles \label{sec:strong}}

In this section we consider the case where we have one type of sources, but the source can produce particles of two distinguishable types.
In this context, the number of fluctuating sources has frequently been referred to as {\em volume fluctuations}. In measuring a
fluctuating quantity, we obviously want to be insensitive to these spurious fluctuations, and separate them from a {\em true} 
correlation mechanism,. 

Let the multiplicities of the two types of produced particles be
\begin{eqnarray}
N_a=\sum_{j=1}^{S} n_{a,j}, \;\; N_b=\sum_{j=1}^{S} n_{b,j},  \label{eq:Nab} 
\end{eqnarray}
where $a$ and $b$ label the particle types. The particles emitted from the same source are, in general, correlated, but as before, particles emitted form 
different sources are uncorrelated. Then, repeating the derivation of the previous sections, we readily find 
\begin{eqnarray}
K^{N_a,N_b}(t_a,t_b)=K^{S}[K^{n_a,n_b}(t_a,t_b)],
\end{eqnarray}
and, for the more general composition of the generating functions, in analogy to Table~\ref{tab:comp},
\begin{eqnarray}
P^{N_a,N_b}(t_a,t_b)=Q^{S}[R^{n_a,n_b}(t_a,t_b)]. \label{eq:2p}
\end{eqnarray}
In the present case the interesting inverse problem concerns the production mechanism from the sources,
\begin{eqnarray}
R^{n_a,n_b}(t_a,t_b)=(Q^{S})^{-1}[P^{N_a,N_b}(t_a,t_b)].
\end{eqnarray}

Interestingly, the problems stated in  this Section, and in particular Eq.~(\ref{eq:2p}), are related to the construction of
the {\em strongly intensive} measures~\cite{Gazdzicki:1992ri,Gorenstein:2011vq,Gazdzicki:2013ana,Sangaline:2015bma} of the event-by-event fluctuations, {\em i.e.}, combinations
of moments of $n_a$ and $n_b$ expressed via moments of $N_a$ and  $N_b$ in a form-invariant way where the moments of 
the sources $S$ do not appear.  In App.~\ref{app:intense} we show how to derive the following combinations of moments: 
the obvious one
\begin{eqnarray}
\frac{P_{01}}{P_{10}} = \frac{R_{01}}{R_{10}},
\end{eqnarray}
two rank-2 relations
\begin{eqnarray}
Q_1 \left ( \frac{P_{20}}{P_{10}^2} - \frac{P_{02}}{P_{01}^2}  \right ) &=&\frac{R_{20}}{R_{10}^2} - \frac{R_{02}}{R_{01}^2} , \label{eq:r2} \\
Q_1 \left ( \frac{P_{20}}{P_{10}^2} - \frac{2P_{11}}{P_{10}P_{01}}+ \frac{P_{02}}{P_{01}^2}  \right ) 
&=&\frac{R_{20}}{R_{10}^2} - \frac{2R_{11}}{R_{10}R_{01}}+ \frac{R_{02}}{R_{01}^2} , \nonumber
\end{eqnarray}
and one rank-3 relation 
\begin{eqnarray}
&& Q_1^2 \left ( \frac{P_{30}}{P_{10}^3} - \frac{3P_{21}}{P_{10}^2P_{01}} + \frac{3P_{12}}{P_{10}P_{01}^2}- \frac{P_{03}}{P_{01}^3}  \right ) 
\nonumber \\ && ~~ = \frac{R_{30}}{R_{10}^3} - \frac{3R_{21}}{R_{10}^2R_{01}} + \frac{3R_{12}}{R_{10}R_{01}^2}- \frac{R_{03}}{R_{01}^3} . \nonumber \\ \label{eq:r3}
\end{eqnarray}
A non-trivial feature of the above formulas, crucial for the application,  is that the structural form of the $R$ and $P$ moments appearing on both sides of the equalities 
is exactly the same. 

One can use Eq.~(\ref{eq:Q1}) to rewrite Eq.~(\ref{eq:r2},\ref{eq:r3}) in an alternative form
\begin{eqnarray}
&& \frac{1}{Q_1} \left ( \frac{P_{01}P_{20}}{P_{10}} - \frac{P_{10}P_{02}}{P_{01}}  \right ) 
=\frac{R_{01}R_{20}}{R_{10}} - \frac{R_{10}R_{02}}{R_{01}} , \label{eq:r2a} \\
&& \frac{1}{Q_1} \left ( \frac{P_{01}P_{20}}{P_{10}} -2P_{11}+ \frac{P_{10}P_{02}}{P_{01}}  \right ) \nonumber \\
&& ~~ = \frac{R_{01}R_{20}}{R_{10}} -2R_{11}+\frac{R_{10}R_{02}}{R_{01}} , \nonumber
\end{eqnarray}
and the rank-3 equation as
\begin{eqnarray}
&& ~~~ \frac{P_{01}P_{30}}{P_{10}^2} - \frac{3P_{21}}{P_{10}} + \frac{3P_{12}}{P_{01}}- \frac{P_{10}P_{03}}{P_{01}^2}  \nonumber \\
&& = \frac{R_{01}R_{30}}{R_{10}^2} - \frac{3R_{21}}{R_{10}} + \frac{3R_{12}}{R_{01}}- \frac{R_{10}R_{03}}{R_{01}^2} . \nonumber \\ \label{eq:r3a}
\end{eqnarray}
In Eq.~(\ref{eq:r2a}) we readily recognize the $\Sigma$ and $\Delta$ measures introduced in~\cite{Gorenstein:2011vq}. The appearance of $Q_1$ my be
canceled by a multiplication or Eqs.~(\ref{eq:r2a}) side-by-side
with the equation 
\begin{eqnarray}
Q_1/(P_{10}\pm P_{01})=1/(R_{10}\pm R_{01}), \label{eq:e1}
\end{eqnarray}
or
\begin{eqnarray}
Q_1/\sqrt{P_{10}P_{01}}=1/\sqrt{R_{10}R_{01}}, \label{eq:e2}
\end{eqnarray}
or, in general, an equation of this form involving $Q_1$ and any intensive 
quantities for the $P$ and $R$ moments~\cite{Gazdzicki:2013ana}.

We note that a form analogous to Eq.~(\ref{eq:r3a}) was proposed a long time ago in~\cite{Mrowczynski:1999un}
as a rank-3 generalization of the $\Phi_{p_T}$ measure used  for the rank-2 moments~\cite{Gazdzicki:1992ri}.

Finally, we note that Eqs.~(\ref{eq:r2},\ref{eq:r3}) can be stated in a more compact form for the case of cumulant moments of 
the scaled numbers of particles, $\hat{N}_i=N_i/\langle N_i \rangle$, and the scaled moments of particles produced from a source, $\hat{n}_i=n_i/\langle n_i \rangle$:
\begin{eqnarray}
Q_1        \left[ \kappa_2(\hat{N}_a )-\kappa_2( \hat{N}_b) \right]&=& \kappa_2(\hat{n}_a )-\kappa_2( \hat{n}_b) , \label{eq:ints}\\
Q_1       \kappa_2( \hat{N}_a- \hat{N}_b )&=&  \kappa_2\left( \hat{n}_a- \hat{n}_b \right), \nonumber \\
Q_1^2  \kappa_3( \hat{N}_a- \hat{N}_b )&=&  \kappa_3\left( \hat{n}_a- \hat{n}_b \right), \nonumber 
\end{eqnarray}
which can be verified explicitly. Higher-rank relations of this type are obtained in Sec.~\ref{sec:dp}.

\section{Cumulants for differences of particle species and new strongly intensive measures \label{sec:dp}} 

As suggested by the simplicity of Eq.~(\ref{eq:ints}), we now consider in a greater detail the scaled moments of the difference of particles of type 
$a$ and $b$ produced from a single type of sources, 
\begin{eqnarray}
 \hat{N}_-= \hat{N}_a- \hat{N}_b, \;\;\;  \hat{n}_-=\hat{n}_a- \hat{n}_b. 
\end{eqnarray}
The cumulant generating function satisfies the composition law
\begin{eqnarray}
&& e^{K^{\hat{N}_-}(t)}=\E e^{t \frac{\langle n_a \rangle}{\langle N_a \rangle} \sum_{i=1}^S \hat{n}_a-
 t \frac{\langle n_b \rangle}{\langle N_b \rangle} \sum_{i=1}^S \hat{n}_b}  \\
&& ~~ =\E e^{\frac{t}{\langle S \rangle} \sum_{i=1}^S(\hat{n}_a-\hat{n}_b)}
=  e^{K^S\left [ K^{\hat{n}_-}(t/\langle S \rangle) \right ]}, \nonumber \
\end{eqnarray}
where we have used the fact that $\langle N_i \rangle = \langle S \rangle \langle n_i \rangle$.
In our generic notation 
\begin{eqnarray}
 P^{\hat{N}_-}(t) =  Q^S\left [ R^{\hat{n}_-}(t/Q_1) \right ],
\end{eqnarray} 
thus we recover the structure of the composition law for the univariate case of Sec.~\ref{sec:onevar}.

Since the variable $t$ is rescaled, in the formulas given this section we have $R$ corresponding to the cumulant moments, whereas $P$ and $Q$ relate to the cumulant or standard moments (cf. rows 2 and 3 of Table~\ref{tab:comp}). One may then always pass to the desired type of moments according to the scheme of Fig.~\ref{fig:tr}.

The use of scaled variables leads to simplification, as
from construction, for the difference of the scaled moments we have 
\begin{equation}
R_1 \equiv \left \langle \frac{N_a}{\langle N_a \rangle} -  \frac{N_b}{\langle N_b \rangle}  \right \rangle=0. 
\end{equation}
In consequence, from Eqs.~(\ref{app:one}) we obtain the following hierarchy of equations
\begin{eqnarray}
P_1 &=& 0, \label{eq:ones} \\
Q_1 P_2 &=& R_2, \nonumber \\
Q_1^2 P_3 &=& R_3, \nonumber \\
Q_1^3 P_4 &=& 3 \hat{Q}_2 R_2^2 +R_4, \nonumber \\
Q_1^4 P_5 &=& 10 \hat{Q}_2 R_2 R_3+R_5, \nonumber \\ 
Q_1^5 P_6 &=& 15 \hat{Q}_3 R_2^3+\hat{Q}_2 \left(10 R_3^2+15 R_2 R_4 \right)+ R_6, \nonumber \\
Q_1^6 P_7 &=& 105 \hat{Q}_3 R_2^2 R_3+\hat{Q}_2 \left(35 R_3 R_4+21 R_2 R_5\right)+ R_7, \nonumber \\
&\dots&, \nonumber
\end{eqnarray}
where for brevity we also have introduced the scaled moments of the sources
\begin{eqnarray}
\hat{Q}_n=Q_n/Q_1, \;\;\;n=2,3,\dots.
\end{eqnarray}
We notice that the second and third equality in (\ref{eq:ones}) are the same as the corresponding equalities in Eq.~(\ref{eq:ints}): the second one is the $\Delta$ 
measure~\cite{Gorenstein:2011vq}, and the third one is its generalization to rank 3, analogous to the relation derived in~\cite{Mrowczynski:1999un} for $\Phi_{p_T}$. 

We now pass to deriving new strongly intensive fluctuation measures.
Eliminating $\hat{Q}_2$ form the 
fourth and fifth equations, which is possible when $R_3 \neq 0$, we arrive at the relation
\begin{eqnarray}
Q_1 \left [\frac{P_4}{3 P_2^2}-\frac{P_5}{10 P_2 P_3} \right] = \frac{R_4}{3 R_2^2}-\frac{R_5}{10 R_2 R_3}, \label{eq:new1}
\end{eqnarray}
involving moments up to rank 5. The next order relation comes via elimination of  $\hat{Q}_3$ from the 
sixth and seventh equalities in Eqs.~(\ref{eq:ones}), and then eliminating  $\hat{Q}_2$ using the fourth equality. 
This requires $Q_5 \neq 0$ and $Q_3 \neq 0$. The result is
\begin{eqnarray}
&& Q_1\! \left [ \frac{7 P_3 P_6- P_2 P_7}{70 P_3^3\!+\!70 P_2 P_4 P_3\!-\!21 P_2^2 P_5}\!-\!(1\!-\!a)\frac{P_4}{3 P_2^2}\!-\!a\frac{P_5}{10 P_2 P_3}\right ] \nonumber \\
&& = \frac{7 R_3 R_6- R_2 R_7}{70 R_3^3\!+\!70 R_2 R_4 R_3\!-\!21 R_2^2 R_5}\!-\!(1\!-\!a)\frac{R_4}{3 R_2^2}\!-\!a\frac{R_5}{10 R_2 R_3}, \nonumber \\  \label{eq:new2}
\end{eqnarray}
where $a$ is any real parameter; the form is not unique, as we may use Eq.~(\ref{eq:new1}) to alter 
the coefficients in front of the terms involving $P_4$ and  $P_5$, or $R_4$ and  $R_5$, respectively. 

The procedure may possibly be continued to yet higher orders, but it becomes tedious (see the discussion in App.~\ref{app:intense}). We should also bear in mind that potential practical significance 
of the formulas decreases with the degree of complication and the increasing rank, as higher order moments are 
subject to larger experimental uncertainties. Present analyses use moments up to rank 4, which is sufficient to apply the second and third equality in Eq.~(\ref{eq:ones}).
A usage of Eqs.~(\ref{eq:new1},\ref{eq:new2}) would require going up to rank 5 and 7, respectively, for which a very large data statistics would be 
necessary.

Division of Eq.~(\ref{eq:new1},\ref{eq:new2}) with Eqs.~(\ref{eq:e1},\ref{eq:e2}), or any equation dependent on $Q_1$
in a similar way, removes the dependence on $Q_1$, in full analogy to the construction of the $\Sigma$ and $\Delta$~\cite{Gorenstein:2011vq} measures. 

As mentioned in the Introduction, important physical applications of moments of differences of particles produced 
in ultra-relativistic heavy-ion collisions, or the corresponding strongly intensive measures, are linked to the quest of the QCD phase 
transition at a finite baryon density. 

\section{Conclusion}

We have reviewed the formalism of generating functions in the application to superposition models used in 
production of particles in hadronic or nuclear high-energy collisions. We have indicated that simple composition laws 
hold for appropriate types of functions (for the standard, cumulant, factorial, and factorial cumulant moments), which allows us 
for a simple derivation of the composition laws for the moments themselves to any order. We have recalled the 
exact transformations between various types of moments, as well as provided the inverse transformation, {\em i.e.}, obtaining the moments of sources from the 
moments of particles. We have drawn attention to the fact that the composition laws hold for numerous combinations of types of moments, as summarized in 
Table~\ref{tab:comp}.

We have then considered the following simple cases: 1)~two 
kinds of sources and a single particle type (e.g., wounded nucleon/quark model, or correlations between multiplicities in 
different kinematic bins), and 2)~one type of sources and two kinds of particles (or two kinds of characteristics, such as multiplicity, charge, or transverse momentum). 
This case arises, e.g., when one considers the net baryon number or charge (the alleged probes of the QCD phase transition). 

A generalization of the master composition formula for the generating functions for more types of sources and particles is 
straightforward
\begin{eqnarray}
&& P^{N_a,N_b,\dots}(t_a,t_b,\dots)= \label{eq:ggen} \\ 
&& Q^{S_A,S_B,\dots}[R^{n^A_a, n^A_b,\dots}(t_{a},t_{b},\dots), R^{n^B_a,n^B_b,\dots}(t_{a},t_{b},\dots), \dots],\nonumber
\end{eqnarray}
where $n^I_j$ is the distribution of particles of type $j$ produced from the source of type $I$.  The corresponding formulas for the composition of moments can be obtained via the
Maclaurin expansion of Eq.~(\ref{eq:ggen}). 

Moreover, the quantities used in our statistical study need not be multiplicities themselves, as used throughout the paper
for the simplicity of notation, but any additive one-body observable, for instance charge or the transverse momentum.
We note that the correlations of the transverse momenta and multiplicities are actively pursued experimentally 
(see, e.g.,~\cite{Aduszkiewicz:2015jna}) with the use of the strongly-intensive measures.

We have used the framework to consider the scaled moments of the difference of multiplicities of two kinds of particles and found 
a straightforward derivation and a simple algebraic interpretation 
of the strongly intensive fluctuation measures. With this method we have derived new relations of that type, relating moments of higher rank. 
Hopefully, these relations can be applied to high statistics data samples, thus will become useful in analyses of the event-by-event fluctuations. 

The generalization of the superposition framework to more types of sources, as in Eq.~(\ref{eq:ggen}), allows for 
incorporation of global conservation laws. The conserved quantity (e.g., charge) may be distributed over the sources of different 
type (labeled with the carried charge). As the total charge of all sources is constrained, the distribution of sources will reflect this, 
which will show up in the $Q$ moments. The production from a source should also be conserving, which 
will affect in the $R$ moments. Nevertheless, the generic structure of Eq.~(\ref{eq:ggen}) remains valid. In the case where there are more particle types
than source types, one would get an over-determined system of equations in analogy to Eqs.~(\ref{eq:Q4}), and 
relations between the particle moments  $P$ and source moments $R$, without a reference to $Q$ moments, may be obtained along the lines of
App.~\ref{app:intense}. 

\begin{acknowledgments}
WB is grateful to Stanis\l{}aw Mr\'owczy\'nski for helpful discussions concerning the strongly intensive measures.
This research was supported by the Polish National Science Centre grant 2015/19/B/ST2/00937.
\end{acknowledgments}
\appendix

\section{Explicit formulas \label{app:explicit}}

In this Appendix we give a glossary of formulas following from the composition laws discussed in the body of the paper for the case of a single type of sources and a single type of the 
produced particles. These formulas can be useful 
in analyses in the superposition approach. For the first few values of $m$ one finds directly from Eq.~(\ref{eq:bell}):
\begin{eqnarray}
P_1 &=& Q_1 R_1, \label{app:one} \\
P_2 &=& Q_2 R_1^2+Q_1 R_2, \nonumber \\
P_3 &=& Q_3 R_1^3+3 Q_2 R_2 R_1+Q_1 R_3, \nonumber \\
P_4 &=& Q_4 R_1^4+6 Q_3 R_2 R_1^2+Q_2 \left(3 R_2^2+4 R_1 R_3\right)+Q_1 R_4, \nonumber \\
P_5 &=& Q_5 R_1^5+10 Q_4 R_2 R_1^3 +Q_3 \left(10 R_3 R_1^2+15 R_2^2 R_1\right) \nonumber \\ 
      &&+ Q_2 \left(10 R_2 R_3+5 R_1 R_4\right)+Q_1 R_5, \nonumber \\ 
P_6 &=& Q_6 R_1^6+15 Q_5 R_2 R_1^4+Q_4 \left(20 R_3 R_1^3+45 R_2^2 R_1^2\right) \nonumber \\
      &&+Q_3 \left(15 R_2^3+60 R_1 R_3 R_2+15 R_1^2 R_4\right) \nonumber \\ 
      &&+Q_2 \left(10 R_3^2+15 R_2 R_4+6 R_1 R_5\right)+Q_1 R_6, \nonumber \\ && \dots \nonumber
%
\end{eqnarray}

For the case of the inverse problem (\ref{eq:invcomp}), the first few terms are 
\begin{eqnarray}
R_1 Q_1 &=& P_1, \label{app:two} \\
R_1^3 Q_2 &=&  2 P_2 R_1-P_1 R_2  ,\nonumber \\
R_1^5 Q_3 &=&  6 P_3 R_1^2-6 P_2 R_2 R_1+P_1 \left(3 R_2^2-R_1 R_3\right)  ,\nonumber \\
R_1^7 Q_4 &=&  24 P_4 R_1^3-36 P_3 R_2 R_1^2+P_2 \left(30 R_1 R_2^2-8 R_1^2 R_3\right) \nonumber \\ 
                  &&+P_1 \left(-15 R_2^3+10 R_1 R_3 R_2-R_1^2 R_4\right)   ,\nonumber \\ && \dots \nonumber
\end{eqnarray}

\section{Derivation of intensive measures\label{app:intense}}

In this Appendix we list the explicit formulas for the case of a single type of sources and two kinds of the produced 
particles. 
From the form of Eqs.~(\ref{eq:2p}) one can, via the Maclaurin expansion, obtain the following hierarchy of equations:
\begin{eqnarray}
P_{01}=Q_1 R_{01}, \;\;\; P_{10}=Q_1 R_{10},  \label{eq:Q1}
\end{eqnarray}

\begin{eqnarray}
P_{20} &= & Q_2 R_{10}^2+Q_1 R_{20}, \nonumber \\
P_{11} &= & Q_2 R_{10}R_{01}+Q_1 R_{11}, \nonumber \\
P_{02} &= & Q_2 R_{01}^2+Q_1 R_{02}, \label{eq:Q2}
\end{eqnarray}

\begin{eqnarray}
P_{30} &= & Q_3 R_{10}^3+3Q_2 R_{10}R_{20}+Q_1 R_{30}, \nonumber \\
P_{21} &=&  Q_3 R_{10}^2 R_{01}+Q_2 \left ( 2R_{10}R_{11}+R_{20}R_{01} \right ) +Q_1 R_{21}, \nonumber \\
P_{12} &=&  Q_3 R_{01}^2 R_{10}+Q_2 \left ( 2R_{01}R_{11}+R_{02}R_{10} \right ) +Q_1 R_{12}, \nonumber \\
P_{03} &= & Q_3 R_{01}^3+3Q_2 R_{01}R_{02}+Q_1 R_{03},  \label{eq:Q3}
\end{eqnarray}

\begin{eqnarray}
P_{40} &=& Q_4 R_{10}^4+6 Q_3 R_{20} R_{10}^2+Q_2 \left(3 R_{20}^2+4 R_{10} R_{30}\right)\nonumber \\ && +Q_1 R_{40}, \nonumber \\
P_{31} &=& Q_4 R_{01} R_{10}^3+Q_3 \left(3 R_{11} R_{10}^2+3 R_{01} R_{20} R_{10}\right) \nonumber \\ 
&& + Q_2 \left(3 R_{11} R_{20}+3 R_{10} R_{21}+R_{01} R_{30}\right)+Q_1 R_{31},\nonumber \\
P_{22} &=&  Q_4 R_{01}^2 R_{10}^2+Q_3 \left(R_{20} R_{01}^2\!+\!4 R_{10} R_{11} R_{01}\!+\!R_{02} R_{10}^2\right) \nonumber  \\ 
&&+Q_2 \left(2 R_{11}^2+2 R_{10} R_{12}+R_{02} R_{20}+2 R_{01} R_{21}\right)\nonumber \\ &&+Q_1 R_{22}, \nonumber \\
P_{13} &=& Q_4 R_{10} R_{01}^3+Q_3 \left(3 R_{11} R_{01}^2+3 R_{02} R_{10} R_{01}\right) \nonumber \\ 
&&+Q_2 \left(R_{03} R_{10}+3 R_{02} R_{11}+3 R_{01} R_{12}\right)+Q_1 R_{13}, \nonumber \\
P_{04} &=& Q_4 R_{01}^4+6 Q_3 R_{02} R_{01}^2+Q_2 \left(3 R_{02}^2+4 R_{01} R_{03}\right)\nonumber \\ &&+Q_1 R_{04}. \label{eq:Q4}
\end{eqnarray}
{\em etc.} Algebraically, the above equations can be viewed as an over-determined set of equations for the variables $Q_i$, thus one can find conditions for existence of a solution. 
From the Rouch\'e -- Capelli theorem it is clear, that Eq.~(\ref{eq:Q2}) leads to 2 conditions between the $P$ and $R$ moments, as well as $Q_1$, Eq.~(\ref{eq:Q3}) leads to 3 
additional conditions, and so on. However, we are seeking the conditions that can be written with the same structural forms for the $P$ moments as for the $R$ moments, 
for instance as in Eqs.~(\ref{eq:r2},\ref{eq:r3}). The problem is tedious for higher rank cases and we were not able to settle it down in general. 

The lowest-rank relations are, however, straightforward to obtain.
Eliminating $Q_2$, $Q_3$, \dots, from Eq.~(\ref{eq:r2},\ref{eq:r3}) one 
finds the desired combinations. For instance, eliminating $Q_2$ form the first and third Eq.~(\ref{eq:Q2}) one arrives at 
the first formula in Eq.~(\ref{eq:r2}). 

We note that the derivation of the form of the 
strongly intensive measures is simpler along the lines of Sec.~\ref{sec:dp}, where we use the generating function for the difference of scaled numbers 
of particles. However,  in that case the problem of arriving at the same structural forms of the $P$ and $R$ moment combinations also becomes algebraically complicated at higher rank.

\bibliography{hydr}

\end{document}